\documentclass[twocolumn,aps,prl,reprint]{revtex4-2}

\usepackage[utf8]{inputenc}
\usepackage{epsf,graphicx}
\usepackage{natbib}
\usepackage{multirow}
\usepackage{amsmath,gensymb,amssymb}
\usepackage{siunitx}
\usepackage[bottom]{footmisc}
\usepackage{lipsum}
\usepackage{dcolumn}% Align table columns on decimal point
\usepackage{xcolor}
\usepackage{upgreek}
\usepackage{float,ulem}

\usepackage{tikz}
\DeclareRobustCommand{\hexagon}{%
{\begin{tikzpicture}[scale=0.1]
\draw (0,0) -- (1,0) -- (1.5,0.866) -- (1,1.732) -- (0,1.732) -- (-0.5,0.866) -- cycle;
\end{tikzpicture}}
}
\DeclareRobustCommand{\pentagon}{%
{\begin{tikzpicture}[scale=0.1]
\draw (90:1) -- (162:1) -- (234:1) -- (306:1) -- (18:1) -- cycle;
\end{tikzpicture}}
}

\usepackage{bm}% bold math
\usepackage{hyperref}
\hypersetup{
colorlinks=true, %colorise les liens
citecolor=magenta,
breaklinks=true, %permet le retour a la ligne dans les liens trop longs
urlcolor= blue, %couleur des hyperliens
linkcolor= blue, %couleur des liens internes
bookmarksopen=true, %si les signets Acrobat sont crees,
}
\usepackage{natbib}
\usepackage{color}
\usepackage[type={CC},modifier={by-nc-sa},version={4.0},]{doclicense}
\definecolor{linkcolor}{rgb}{0,0,0.6}

\begin{document}
\title{Thermodynamics and statistical equilibrium of\\ large-scale hydroelastic wave turbulence}
%\title{Statistical equilibrium of large scales in hydroelastic wave turbulence}
%\title{Thermal equilibrium of the large scales of hydroelastic waves }

\author{Marlone Vernet}
\author{Eric Falcon}%
\email{eric.falcon@u-paris.fr}
%\orcid{0000-0001-9640-9895}
\affiliation{Université Paris Cité, CNRS, MSC, UMR 7057, F-75 013 Paris, France}
%\date{\today} 

\begin{abstract}
Understanding how statistical equilibrium can occur in out-of-equilibrium systems is of paramount interest, as it would enable the use of statistical mechanics tools to these systems. Here, we report the first experimental evidence of statistical equilibrium of the large scales of hydroelastic turbulent waves driven by small-scale random forcing. The wave field statistics at scales larger than the forcing scale, resolved in space and time, align well with the predictions of Rayleigh-Jeans equilibrium spectra over more than a decade. We measure zero net energy flux in this regime, as expected. We also determine the effective temperature, entropy, and heat capacity of this nonequilibrium system, demonstrating that classical thermodynamic concepts apply to describe large scales in statistical equilibrium of turbulent systems.
%We managed to perform both spatial and temporal measurements, rendering conclusive results. 
%The search for a statistical description of out-of-equilibrium systems remains one of the major challenges in physics. 
\end{abstract}

%\keywords{Wave turbulence, large scales, thermal equilibrium, statistical equilibrium}

\maketitle

%Turbulence can be defined as the chaotic dynamics occurring in a system where a scale separation is present between the forcing scale and the small (dissipative) scale. Then, the flux of conserved quantity occurs through nonlinear transfers between modes. This picture has been widely validated in both hydrodynamical turbulence and wave turbulence and succeeded in describing a wide variety of systems~\cite{alexakis2018,falcon2022}. However, the behavior of scales larger than the forcing scale (in the absence of an inverse cascade) remains partially understood and lacks experimental investigation. The possibility for the large scales of hydrodynamical turbulence to be organized by the energy equipartition principle traces back to the pioneering work of Burgers, Hopf, Lee, Batchelor, and Kraichnan~\cite{lee1952,hopf1952,batchelor1956,kraichnan1975,kraichnan1975JFM}. Such regime has been reported in several numeral studies~\cite{frisch2008,dallas2015,alexakis2019}. While predicted many decades ago, the experimental evidence of the statistical equilibrium, also called ``thermal'' equilibrium, for the large scales in turbulence is quite recent~\cite{gorce2022PRL,gorce2024PRL}. %In their hydrodynamic experiment, small-scale turbulence was driven by magnetically stirred magnets.   

\textit{Introduction---}Wave turbulence has been extensively investigated to understand energy transfers within the inertial range from the energy injection scale to the small, dissipative scale~\cite{FalconARFM2022}. However, the behavior of scales larger than the forcing scale remains partially understood. It has been conjectured since the 50s that the large-scale modes in three-dimensional (3D) turbulence have the same energy and are thus in a statistically stationary equilibrium state~\cite{alexakis2018,lee1952}. This equipartition regime, also called thermal equilibrium, would arise if no energy flux is transferred from the forcing scale to the large scales. While numerical simulations have confirmed the presence of statistical equilibrium in 3D forced turbulent flows~\cite{frisch2008}, experimental evidence of the statistical equilibrium of the large scales in 3D turbulence has only recently emerged~\cite{gorce2022PRL,gorce2024PRL}.  In wave turbulence, statistical equilibrium has been predicted by weak turbulence theory for nearly all wave systems without inverse cascade toward large scales~\cite{FalconARFM2022,zakharov1992,NazarenkoBook,balkovsky1995PRE}. However, the few studies reporting experimental investigations of statistical equilibrium concern capillary wave turbulence on a fluid surface~\cite{michel2017} and flexural wave turbulence in a thin plate~\cite{miquel2021PRE}, using a single point measurement. Condensation of classical optical waves, from equilibrium statistics to the fundamental mode, has also been reported~\cite{SunNatPhys2012,BaudinPRL2020}. %and applied in nearly all domains % {a set of random nonlinear waves in interaction,}

Here, we report the first experimental evidence of the large-scale statistical equilibrium in hydroelastic wave turbulence, resolved in both time and space over more than a decade. We measure zero net energy flux in this regime. We also extend thermodynamic concepts (entropy, heat capacity, Boltzmann energy distribution) to this nonequilibrium system. This approach is of great interest in various theoretical and numerical contexts, including glasses~\cite{CugliandoloPRE1997}, active matter~\cite{FodorPRL2016}, {3D} turbulence~\cite{RuellePNAS2012}, nonlinear optics~\cite{PicozziPR2014}, and quantum fluids~\cite{ZhuArXiv2024}, but remains, in general, an open challenge.

%hopf1952,batchelor1956,Saffman67,KraichnanJFM1973
%kraichnan1975,kraichnan1975JFM
%,dallas2015,alexakis2019,AlexakisJFM2020

%The experimental study of hydroelastic waves also presents interesting applications on its own as it is a good approximation of waves traveling through large parcels of ice floe at the surface of the ocean, 
Hydroelastic waves are also highly relevant for practical applications of great importance. In oceanography, they provide a good approximation of waves propagating on the ice-covered ocean surface~\cite{Parau2024,schulkes1987}. The recent development of very large floating structures, such as floating airports, mobile offshore bases, and large floating solar-panel farms, further highlights the need for a better understanding of these waves~\cite{yang2024}. In both cases, ocean swell can act as small-scale forcing, leading to complex large-scale dynamics.  At the laboratory scale, experimental studies on hydroelastic waves span various contexts, including wave turbulence (direct cascade toward smaller scales)~\cite{deike2013JFM,deike2017PRF}, elastic floating plates to mimic wave-ice floe interaction~\cite{MontielJFM2013,MeylanPoF2015}, wake phenomenona~\cite{OnoDitBiotPRF2019}, bandgaps similar to those in solid-state physics~\cite{DominoAPL2020}, and sub-wavelength focusing~\cite{DominoEPL2018}.

\textit{Theoretical predictions---}For wave turbulence systems involving three-wave nonlinear interactions, wave action is not conserved, and so there is no inverse cascade carrying action flux toward the large scales~\cite{zakharov1992,NazarenkoBook}. Wave dynamics at scales larger than the forcing scale is hence expected to follow a statistical (or thermodynamic) equilibrium regime, that is, the kinetic energy equipartition among the Fourier modes $\mathbf{k}$~\cite{balkovsky1995PRE}. For  waves propagating in two dimensions, the corresponding theoretical energy spectral density reads~\cite{zakharov1992} (see also Supp. Mat.~\cite{SM})
\begin{equation}
E^{\mathrm{Eq}}(k) = \frac{k_B \theta}{2\pi\rho}k {\mathrm ,}
\label{RJspectrum}
\end{equation}
where $k_B$ is the Boltzmann constant, $\theta$ is an effective temperature and $\rho$ is the fluid density. Equation~\eqref{RJspectrum} is the analog for classical waves of the Rayleigh-Jeans spectrum of the blackbody electromagnetic radiation.

Hydroelastic waves are described by combining the F\"oppl-von K\'arm\'an equations, governing the deformation of a thin elastic sheet, with Bernoulli's theorem for a perfect fluid~\cite{Landau}. Neglecting sheet inertia ($ke \ll \rho/\rho_s$ with $e$ and $\rho_s$ the sheet thickness and density), the  dispersion relation of linear hydroelastic waves, in a deep water regime ($kh\gg 1$ with $h$ the fluid depth), then reads~\cite{schulkes1987,marchenko1991,deike2013JFM}
\begin{equation}
\omega^2 = gk + \frac{T}{\rho}k^3 + \frac{B}{\rho}k^5,
\label{eq:LDR}
\end{equation}
where $\omega=2\pi f$ is the angular frequency and $k$ is the wave-number modulus. The three terms of the right-hand side of Eq.~\eqref{eq:LDR} correspond to gravity waves, tensional waves, and bending (or flexural) waves, respectively, with gravity acceleration $g$, and tension applied to the sheet $T$. The sheet has a bending modulus $B = Ee^3 /[12(1-\nu^2)]$, Young's modulus $E$ and Poisson's ratio $\nu$.

%It has been shown theoretically and experimentally that hydroelastic tensional waves present {three-wave interactions}~\cite{marchenko1987,deike2017PRF,pierce2024}. Thus, the wave action is not conserved, and an inverse cascade should be excluded, leaving space for the statistical equilibrium of the large scales. The Rayleigh-Jeans distribution then gives the {theoretical} energy {spectral density}~\cite{zakharov1992}  %balkovsky1995PRE
%The corresponding power-spectral densities of wave amplitude in frequency, $S^{\mathrm{Eq}}_\eta(\omega)$, and in wave number, $S^{\mathrm{Eq}}_\eta(k)$, are then predicted as follows. 

As hydroelastic waves involve three-wave interactions~\cite{marchenko1987,marchenko1991,deike2017PRF,pierce2024}, their large scales are thus expected to reach a statistical equilibrium state. As justified below, considering only hydroelastic tensional waves in Eq.~\eqref{eq:LDR}, using Eq.~\eqref{RJspectrum} and the link between the energy spectrum and the wave-amplitude power spectrum, $E^{\mathrm{Eq}}(k)=(T/\rho) k^2 S^{\mathrm{Eq}}_\eta(k)$, lead to the power spectrum of hydroelastic tensional waves in a statistical equilibrium state, as
\begin{equation}
S^{\mathrm{Eq}}_\eta(k) = \frac{k_B \theta}{2\pi T}k^{-1} {\mathrm .}
\label{eq:prediction_psd_k}
\end{equation}
As this spatial spectrum is linked to the frequency spectrum by $S_\eta(k)dk = S_\eta(\omega)d\omega$, using the dispersion relation and  Eq.~\eqref{eq:prediction_psd_k}, give the frequency spectrum of statistical equilibrium of hydroelastic tensional waves, as
\begin{equation}
S^{\mathrm{Eq}}_\eta(f) = \frac{k_B\theta}{3\pi T} f^{-1}.
\label{eq:prediction_psd_f}
\end{equation}
To our knowledge, evidencing statistical equilibrium by both the frequency spectrum and the spatial spectrum has never been reported experimentally for any wave turbulence systems.

\begin{figure}[t!]
\includegraphics[width=\columnwidth]{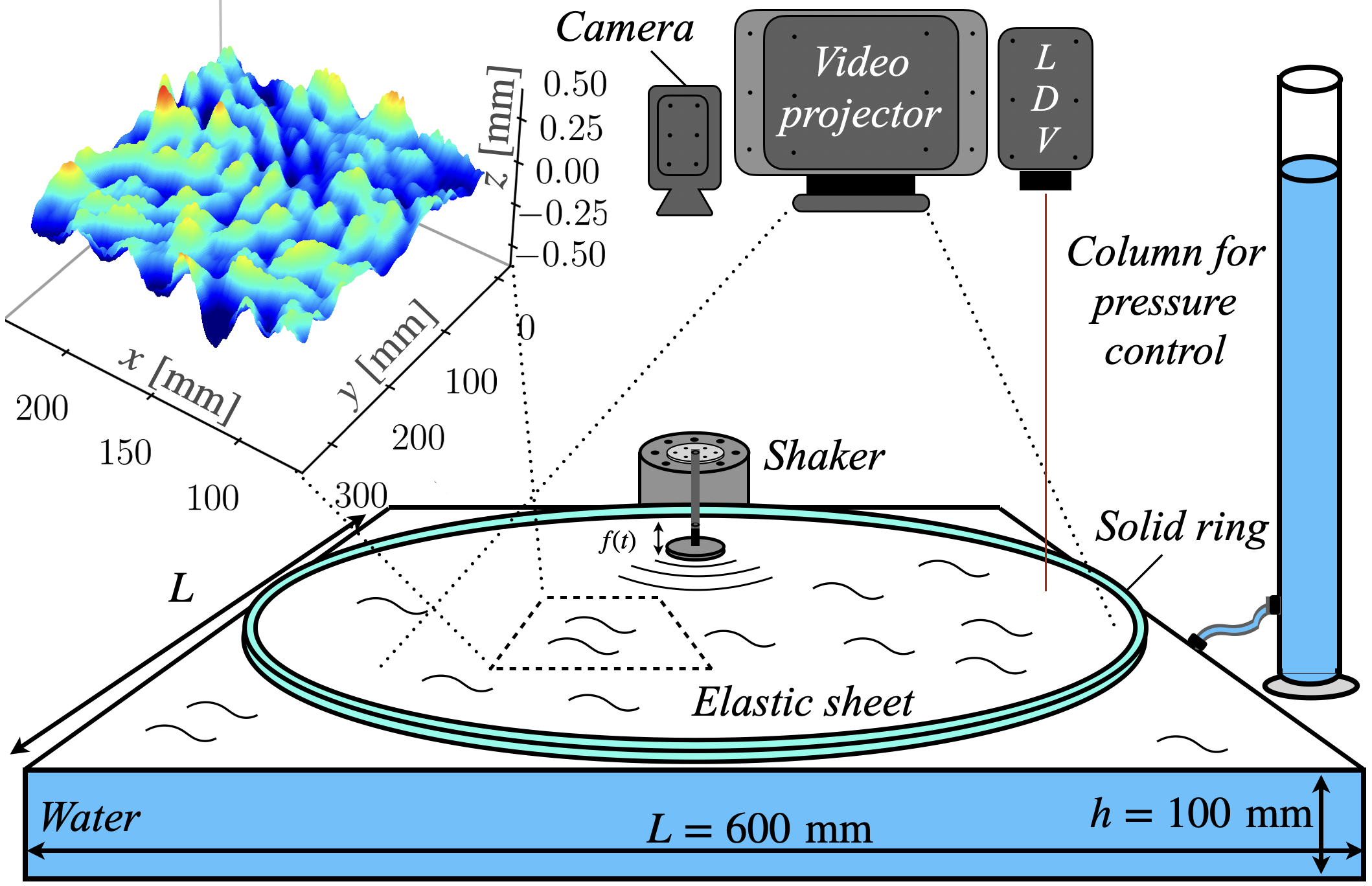}% Here is how to import EPS art
\caption{\label{fig:setup} Scheme of the experimental set-up. Top-left: typical amplitude $\eta(x,y)$ of hydroelastic waves.}
\end{figure}

\textit{Experimental setup---}The experimental setup is shown in Fig.~\ref{fig:setup}. It consists of a square tank of $L\times L \times h$ with a depth $h=100$~mm and $L=600$~mm. The tank is fully filled with water and covered with a white elastic sheet made of silicone rubber (Ecoflex 00-30 soft elastomer) of thickness $e=0.5$~mm, density $\rho_s/ \rho \approx 1$, $E=7\ 10^4$~Pa, and $\nu\approx 0.5$~\cite{DelorySM2024}. A column filled with water is connected to the bath, to control the imposed pressure in the liquid and the mean applied tension to the sheet in the range $T\in [1,7]$~N\ m$^{-1}$. Hydroelastic waves are generated by the vertical motion of a circular wavemaker, $50$~mm in diameter, driven by an electromagnetic shaker (LDS V406), randomly in a frequency range $f_p\in [50,100]$~Hz. The shaker is fed with a bandpass-filtered Gaussian white noise signal. An accelerometer is placed on top of the wavemaker to measure its acceleration and velocity. The vertical deformations of the sheet are either measured at a given position with a laser Doppler velocimeter (LDV) (Polytec OFV5000), or are fully resolved in space and time using the Fourier Transform Profilometry (FTP) method~\cite{cobelli2009} with a camera (Basler acA2040) recording at 120~fps the deformation of a fringe pattern projected over the sheet by a full-HD video projector (Epson EH-TW9400). The acquisition frequency for the accelerometer and the LDV is $2$~kHz. A solid ring placed on the tank plays a crucial role in highlighting statistical equilibrium, as it favors wave reflections in all directions while preventing the emergence of the square tank eigenmodes (see Supp. Mat.~\cite{SM}).  Note that the sheet at rest is flat everywhere except close to the solid ring where all the curvature is confined.%, and the temporal signal is transmitted to a computer through an acquisition card (NI6216) %(monitored by a BK2692 charge amplifier). %{fed by} an AC power supply amplifier (LPA100)

\begin{figure}[t!]
\includegraphics[width=0.9\columnwidth]{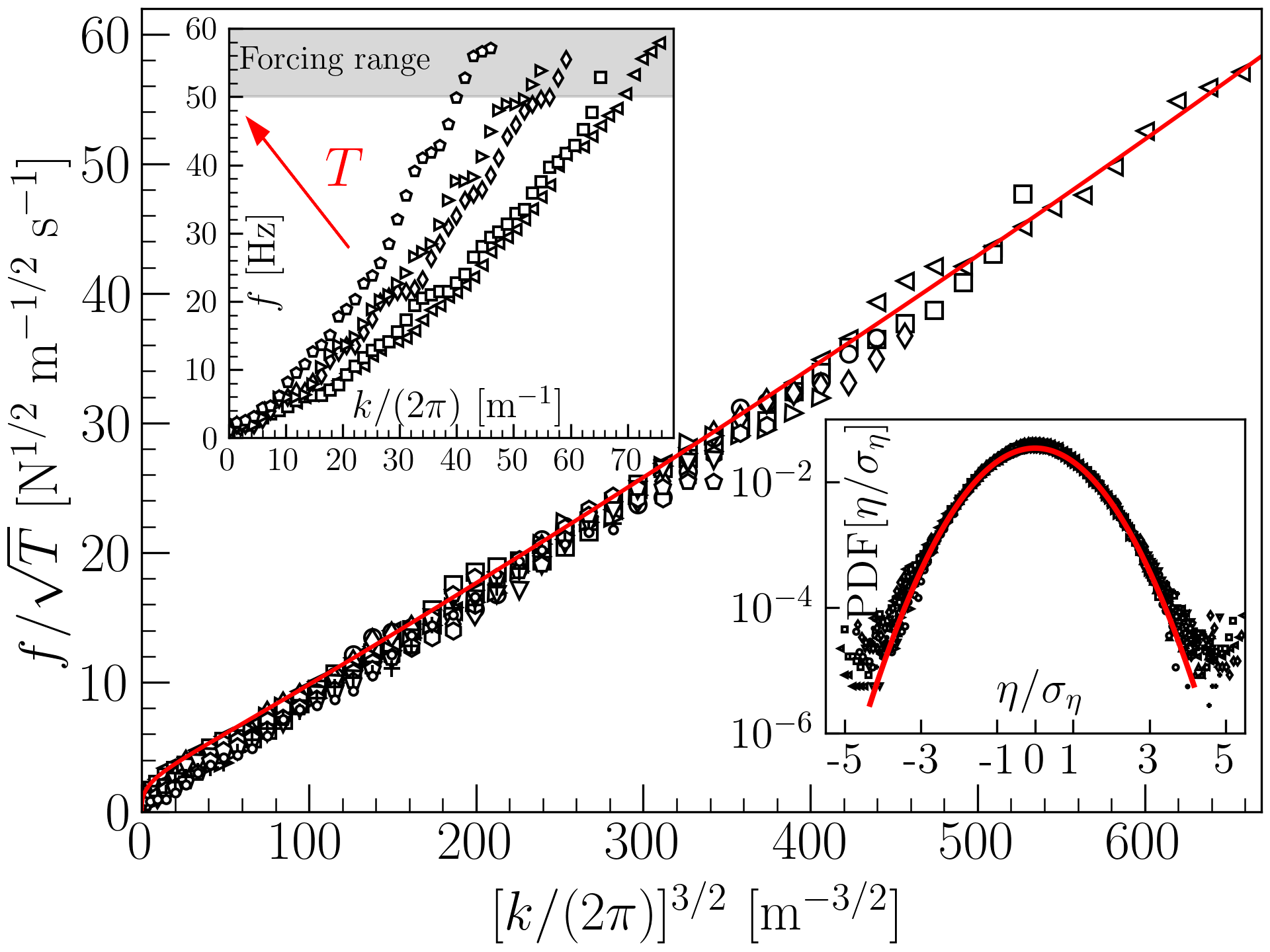}%relation_dispersion_k_VS_LDR_INSET_0711_avgN3_REFEREE.png}% relation_dispersion_k_VS_LDR_INSET_0711_avgN3-1
\caption{\label{fig:dispersion_relation}Rescaled dispersion relation, $f/\sqrt{T}$ vs. $k^{3/2}$, for various applied tensions ${T}$: ($\lhd$) 1.0, ($\Box$) 1.2, ($\triangle$) 1.4, ($\circ$)~1.8, ($\diamond$) 2.3, ($\rhd$) 2.8, (\hexagon ) 3.6, (\pentagon ) 4.1, ($\triangledown$) 5.1,  ($+$) 6.1, ($\circ$) 7.0~N/m. Random forcing $[50,100]$~Hz. Solid line corresponds to Eq.~\eqref{eq:LDR} with no fitting parameter. {Top} inset: Unrescaled dispersion relation for five different tensions $T$ (see arrow). Bottom inset: PDF of the normalized wave amplitude, $\eta/\sqrt{\langle\eta^2\rangle_t}$, for different tensions $T$. The solid line represents a normal distribution. Same symbols as in the main figure.}
\end{figure}

\textit{Dispersion relation---}To investigate the behavior of the large scales of hydroelastic waves, the system is forced at small scales, randomly in a frequency range $f_p$. Random hydroelastic waves are then observed, as shown in Fig.~\ref{fig:setup} (top-left). Typical wave steepness is 0.02, and typical wave amplitude is 1~mm. The spatiotemporal spectrum $S_\eta(k,\omega)$ of the wave field, $\eta(x,y,t)$, is then computed by performing a Fourier transform in space and time, then a polar averaging in Fourier space to obtain $\hat{\eta}(k,\omega)$ and thus $S_\eta(k,\omega)\equiv \vert \hat{\eta}(k,\omega)\vert^2/(L^2\Delta t)$, with $L$ the ring diameter and $\Delta t= 60$~s the acquisition time. The experimental dispersion relation is obtained from the maximum value of the spectrum $S_\eta (\omega, k)$ at each $k$. The inset of Fig.~\ref{fig:dispersion_relation} shows the experimental dispersion relation obtained for different applied tensions $T$. Remarkably, the wave energy, injected at small scales, is spread over all scales larger than the forcing scale. The dispersion relation is steeper when $T$ is higher, as expected from Eq.~\eqref{eq:LDR}. The tension value is inferred by a polynomial fit of data to Eq.~\eqref{eq:LDR}, as $T$ is the only unknown parameter. When rescaled by $1/\sqrt{T}$, all the experimental data collapse to a single curve (see main Fig.~\ref{fig:dispersion_relation}), highlighting the tensional nature of the waves. Note that for the largest scales (lowest $k$) gravity slightly affects waves. Indeed, the typical length scale separating the tensional and gravity regimes in Eq.~\eqref{eq:LDR} is $\ell_{gt} \equiv 2\pi\sqrt{T/\rho g}$ with $1/\ell_{gt}\in[6-16]$~m$^{-1}$. Note that this transition is smooth and below $1/\ell_{gt}$, the tension still affects the waves. Bending waves are here negligible as the transition occurs at $\ell_{tb} \equiv 2\pi\sqrt{B/T}$ with $1/\ell_{tb}\sim 300$~m$^{-1}$, whereas the forcing scale is $1/l_p\sim 50$~m$^{-1}$.

\begin{figure}[t!]
\includegraphics[width=\columnwidth]{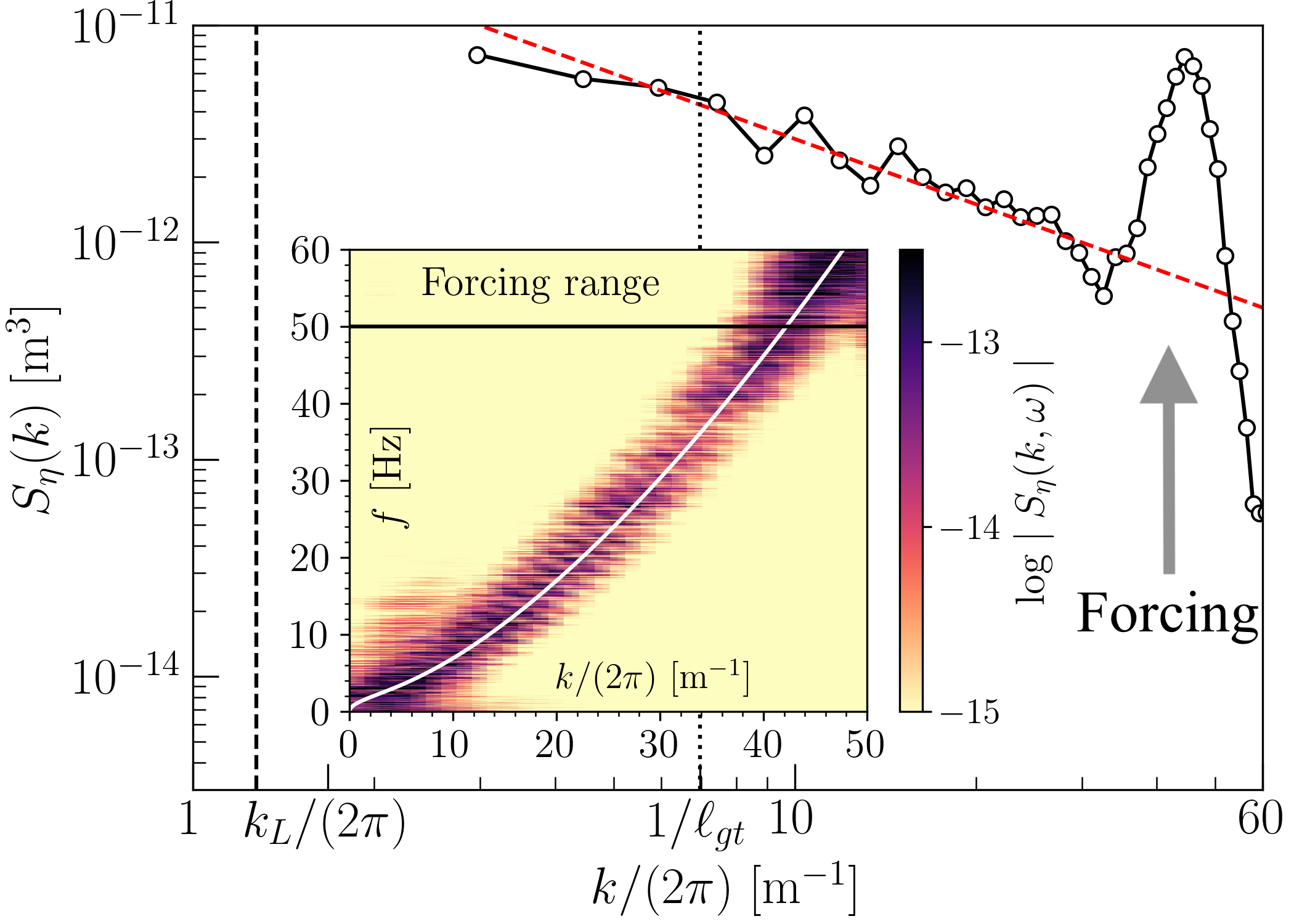}%relation_dispersion_k_50Hz_100Hz_h0_5_T8_PAD2_INSET.png}%PSD_k_50Hz_100Hz_h0_5_T8_PAD2_INSET.png}%PSD_k_50Hz_100Hz_h0_5_T8_PAD2_INSET_V2-1.png}
\caption{\label{fig:psd_k} Power spectral density of the wave amplitude $S_\eta(k)$ as a function of $k/2\pi$ for a tension $T=5.1$~N~m$^{-1}$. The red dashed line has a slope $-1$. The arrow indicates the corresponding forcing range in $k$. Black-dashed line: first axisymetrical eigenmode $k_L/(2\pi)${~\cite{SM,Morse}}. Black-dotted line: $1/\ell_{gt}$. Inset: spatiotemporal spectrum $S_\eta(k,\omega)$ versus $f$ and $k/2\pi$. Solid white line corresponds to Eq.~\eqref{eq:LDR}. Black line indicates the limit above which the forcing occurs.}
\end{figure}

\begin{figure}[t!]
\includegraphics[width=\columnwidth]{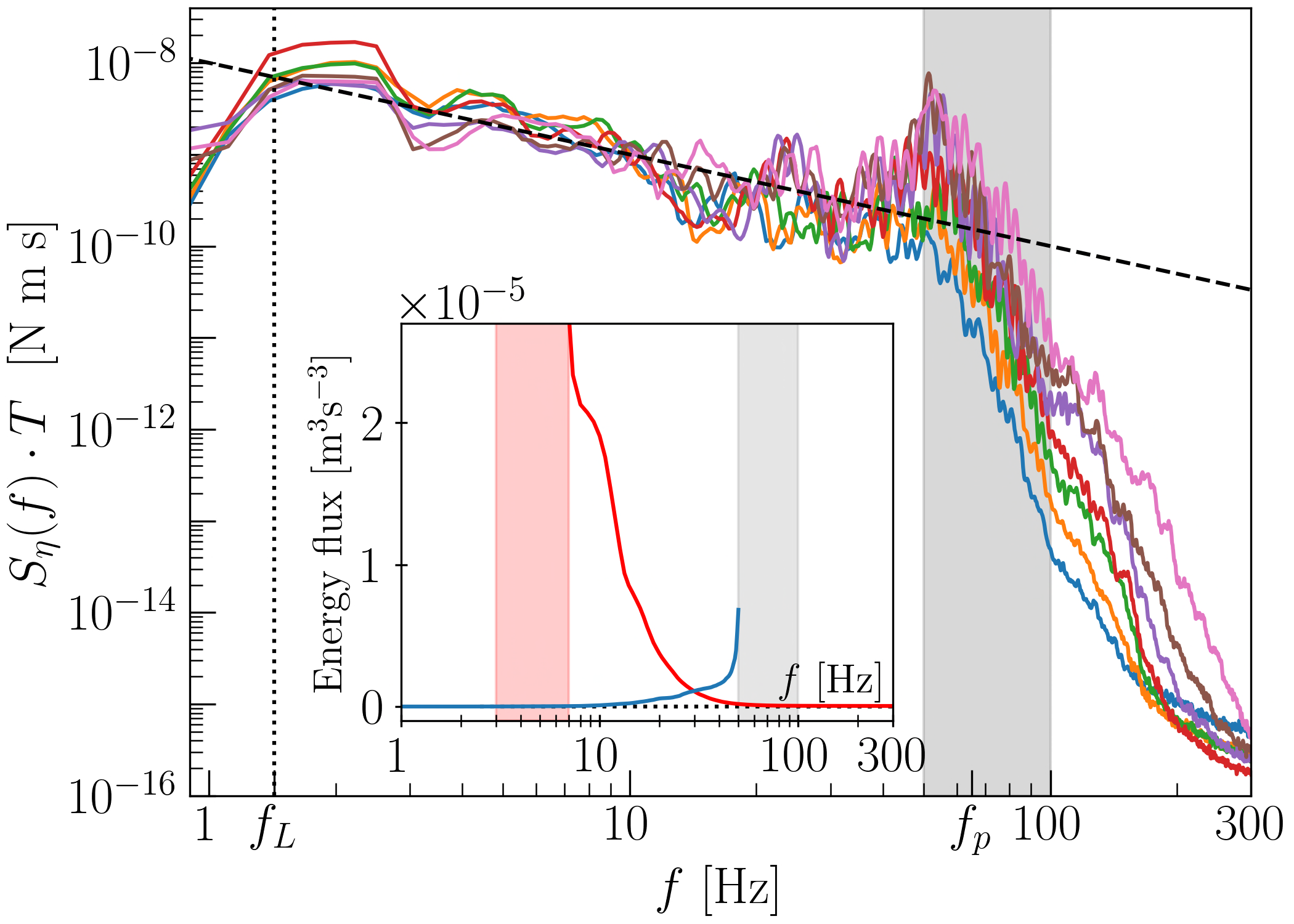}%PDF_vs_Tension_50Hz_100Hz_071124_N5_INSET_DOUBLE_latin}%PDF_vs_Tension_50Hz_100Hz_071124_N5_V2.png}%PDF_vs_Tension_50Hz_100Hz_071124_N5_V2-1.png}
\caption{\label{fig:psd_f}Power spectral density of the wave amplitude $S_\eta(f)$ rescaled by $T$ as a function of the frequency $f$ for various tension values $T$. Dashed line has a slope of $-1$. Dotted vertical line indicates the first axisymmetrical eigenmode $f_L={1.4}$~Hz. Grey region: forcing $f_p\in [50,100]$~Hz. Inset: energy flux $\epsilon(f)$ measured in the statistical equilibrium regime (blue line) for a small-scale forcing (grey region), and in a reference experiment (red line) for a large-scale forcing (red region) of same strength.}
\end{figure}
%At larger $k$, a very steep spectrum is observed due to {small-scale} dissipation, and thus hydroelastic wave turbulence cannot be observed {here}. 

\textit{Equilibrium power spectra---}The spatial power spectrum of the wave amplitude, $S_\eta(k)$ is computed experimentally by averaging the spatiotemporal spectrum $S_\eta(k,\omega)$ over the frequency range. Figure~\ref{fig:psd_k} shows that the spatial spectrum prediction of the statistical equilibrium given by Eq.~\eqref{eq:prediction_psd_k} (see dashed line) is well verified over one decade experimentally. The wide peak at small scales corresponds to the energy injection at $k_p(f_p) \in[35,50]$~m$^{-1}$. The inset of the Fig.~\ref{fig:psd_k} shows the spatiotemporal spectrum $S_\eta(k,\omega)$ for the same experimental data. The energy is indeed concentrated around the theoretical dispersion relation (see white line) of linear waves as expected for the equipartition of the energy among large-scale Fourier modes. The frequency spectrum prediction of Eq.~\eqref{eq:prediction_psd_f} is also well verified experimentally as shown in Fig.~\ref{fig:psd_f} for various tension values. $S_\eta(f)$ is obtained using the LDV measurement at a given position and is indeed found to scale as $f^{-1}/T$ over more than one decade in $f$ from the forcing frequencies $f_p$ down to $f_L$ corresponding to the system's first eigenmode. This is thus a second independent evidence of statistical equilibrium of large-scale hydroelastic waves. Following~\cite{DeikePRE2014}, we measure the experimental energy flux $\epsilon(f)$ from the spectral energy dissipation (see Supp. Mat.~\cite{SM}). As shown in the inset of Fig.~\ref{fig:psd_f}, for far enough forcing scales, we find zero net energy flux within statistical equilibrium (blue curve), as expected. Note that large-scale dissipation is measured to be less than 5\%~\cite{SM}, thus not affecting statistical equilibrium~\cite{LvovEPL2015}. Finally, at small scales ($f>f_p$), no wave turbulence spectrum is observed in main Fig.~\ref{fig:psd_f} as strong dissipation leads to a steep spectrum $\sim f^{-8}$, thus far from previous experimental observations of tensional wave turbulence with a latex sheet~\cite{deike2013JFM}. As expected by statistical equilibrium, the probability density function (PDF) of the wave amplitude, $\eta(t)$, normalized by its rms value $\sigma_{\eta}$, is found to be Gaussian regardless of the applied tension (see inset of Fig.~\ref{fig:dispersion_relation}). Moreover, the experimental PDF of the temporal fluctuations of the wave energy follows an exponential Boltzmann-like distribution (see Supp. Mat.~\cite{SM}). %as expected by equilibrium statistical mechanics

\begin{figure}[t!]
\includegraphics[width=0.9\columnwidth]{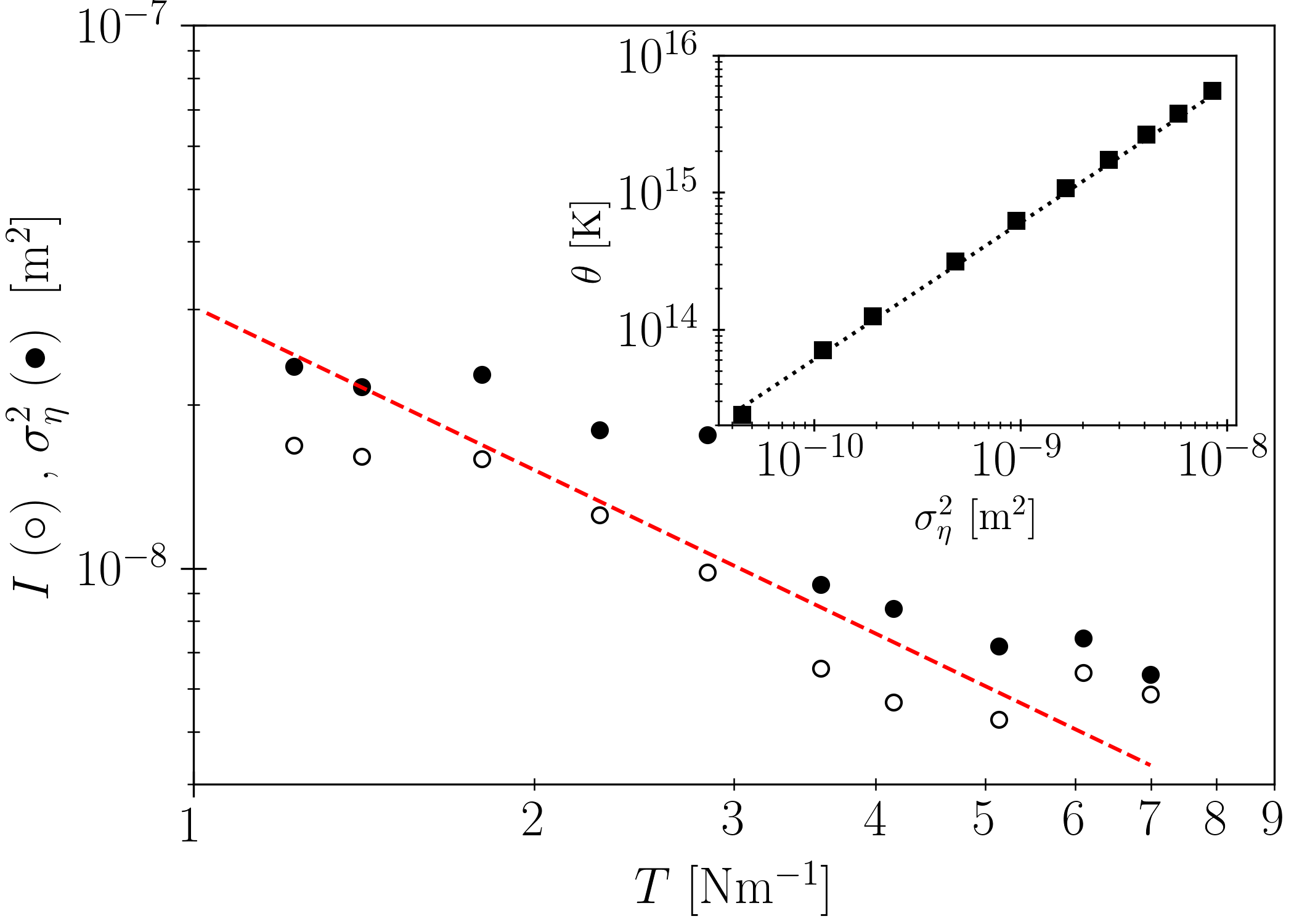}%VARIANCE_ETA_FILTRE_vs_Tension_50Hz_100Hz_071124_INSET-1.png}
\caption{\label{fig:I_vs_T}Effective temperature ${\theta}$ of the statistical equilibrium regime of the large scales, estimated by ($\circ$) the integral $I$ of the frequency spectrum versus the applied tension $T$ and by ($\bullet$) directly computing the wave amplitude variance $\sigma^2_{\eta}$. Red dashed line corresponds to right-hand side of Eq.~\eqref{eq:prediction_I}, with ${\theta}\simeq 8\ 10^{15}$~K. Inset: ${\theta}$ versus $\sigma^2_{\eta}$ for different forcing amplitude at a fixed tension $T=2.8$~N m$^{-1}$.} %$\frac{2k_B T_{\mathrm{eff}}}{3\pi T}\ln{(f_{p}/f_{L})}$ from
\end{figure}

\textit{Effective temperature---}One way to estimate the effective temperature ${\theta}$ is to integrate both members of Eq.~\eqref{eq:prediction_psd_f} over the large-scale frequency range fulfilling the statistical equilibrium prediction, as
\begin{equation}
	I \equiv \int_{f_{L}}^{\min(f_{p})}S_\eta(f)df = \frac{k_B \theta}{3\pi T}\ln{[\min(f_{p})/f_{L}]} .
\label{eq:prediction_I}
\end{equation}
The experimental values of the integral $I$ are shown in Fig.~\ref{fig:I_vs_T} for different $T$ at fixed forcing amplitude. They are roughly found to decrease as the inverse of the applied tension, $1/T$, as predicted by the right-hand side of Eq.~\eqref{eq:prediction_I} shown by the dashed line in Fig.~\ref{fig:I_vs_T}, its slope leading thus a measurement of the effective temperature as ${\theta}\simeq 8~10^{15}$~K. ${\theta}$ is found to be 13 orders of magnitude higher than the room temperature, as also found for statistical equilibrium of large scales of capillary wave turbulence~\cite{michel2017} and of 3D turbulence~\cite{gorce2022PRL}. One can also estimate the effective temperature by directly computing the wave-amplitude variance $\sigma^2_{\eta}=\langle \eta^2\rangle_t$ , which is equal, according to Parseval's theorem, to the area $I$ under the frequency spectrum, and thus to $\sigma^2_{\eta} \sim \theta/T$, using Eq.~\eqref{eq:prediction_I}. These scaling laws are well-verified experimentally in Fig.~\ref{fig:I_vs_T}, this second estimation leading to similar temperature values. In practice, the temporal signal $\eta(t)$ is low-pass filtered below $f_{p}$ when computing $\sigma^2_{\eta}$ to focus only on the large-scale range. Finally, by varying the forcing strength, we find that the experimental effective temperature scales as ${\theta}\sim\sigma_\eta^2$ as shown in the inset of Fig.~\ref{fig:I_vs_T} (see corresponding spectra in Supp. Mat.~\cite{SM}). %Finally, we can also obtain the dependence of the effective temperature of the large scales with forcing strength. To do so, we measure the power injected by the wavemaker, $\mathcal{P} = \langle \mathcal{F}\cdot V \rangle_t$, where $\mathcal{F}$ is the force applied by the shaker to the sheet, and $V$ is the wavemaker velocity. We find experimentally ${\theta}\propto \sigma_V^3$ and $\mathcal{P} \sim \sigma_V^2$ on two decades (see Supp. Mat.~\cite{SM}), where $\sigma_V=\sqrt{\langle V^2\rangle_t}$ is the rms value of $V$. One has thus ${\theta}\propto \mathcal{P}^{3/2}$.

%{\color{blue} Of course, $I$ is related to the standard deviation of the height $\sigma_\eta$ by mean of Parseval's theorem $\sqrt{I}=\sigma_\eta^{<f_c}$. Here, $f_c$ denotes the cutoff frequency of the low-pass filtering applied to the temporal signal $\eta(t)$ with $f_c=f_{inj}$. }

%The effective temperature of the large scales {is} also known to increase as $\epsilon^{1/2}$, $\epsilon$ being the energy flux of the wave turbulence cascade~\cite{balkovsky1995PRE}. Unfortunately, as no wave turbulence is produced at small scales, this prediction could not be tested. 

\textit{Entropy and heat capacity---}The above evidence of a statistical equilibrium regime for large-scale random waves paves the way to apply classical thermodynamic concepts to describe them. 
\begin{figure}
\includegraphics[width=0.9\columnwidth]{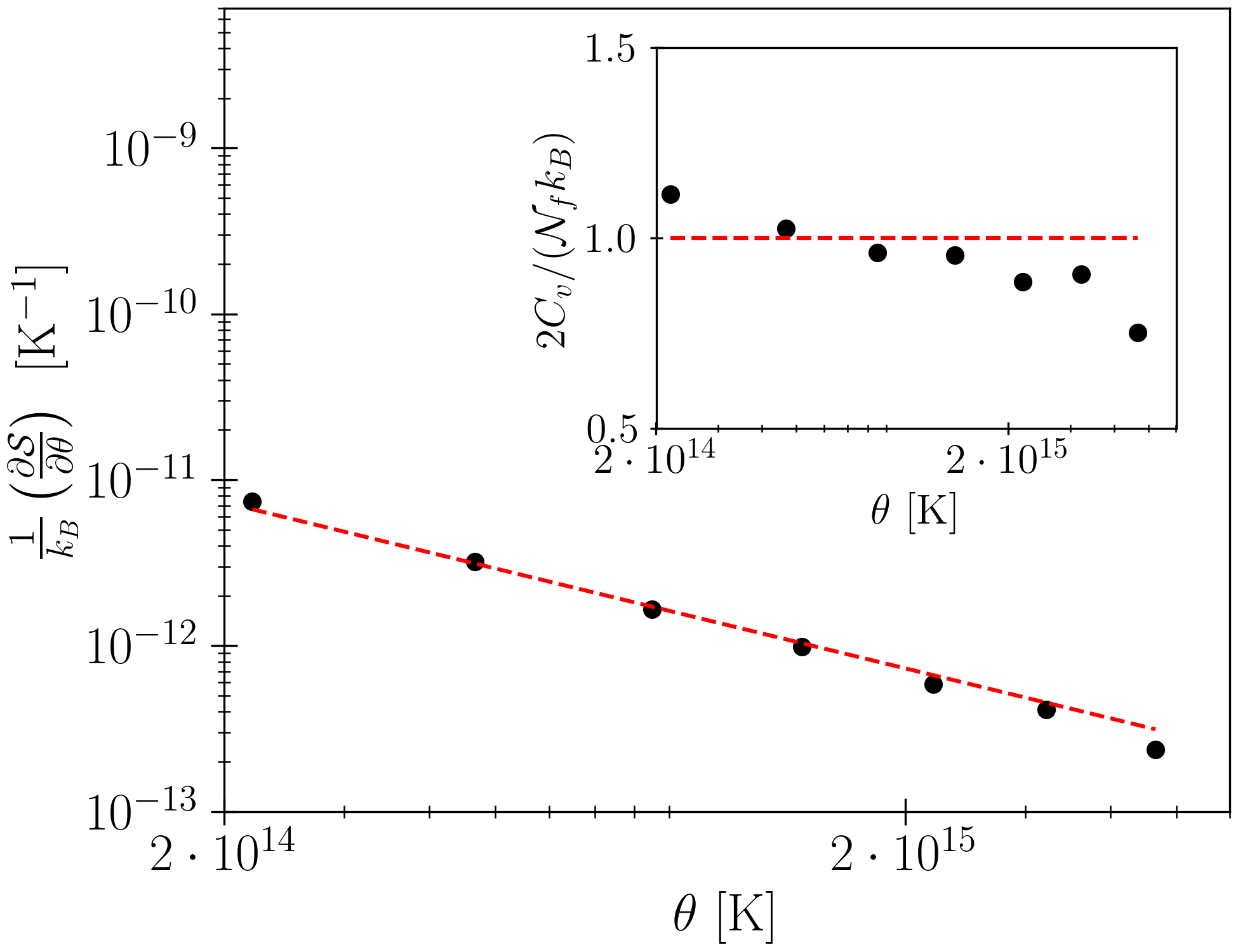}
\caption{\label{fig:S_dT} Evolution of the partial derivative of entropy to~${\theta}$ versus the effective temperature. Different forcing amplitudes. Dashed line has a slope of $-1$. Inset: rescaled experimental heat capacity $C_v$. Dashed line shows the prediction of Eq.~\eqref{eq:heat_cap}.}
\end{figure}
For instance, from $H$ theorem, the statistical equilibrium is known to maximize the entropy ${\mathcal S} \equiv k_B \int \log{[n(\mathbf{k})]d\mathbf{k}} \left(\frac{L}{2\pi}\right)^2$ of the wave system where $n(\mathbf{k})\equiv E(\mathbf{k})/\omega=n(\omega)$ is the wave action~\cite{zakharov1992}. ${\mathcal S}$ can be inferred from the experimental temporal spectrum $S_\eta(\omega)$. Indeed, using the dispersion relation of tensional waves $\omega=\sqrt{T/\rho}k^{3/2}$, the entropy reads ${\mathcal S} = \frac{k_BL^2}{3\pi}(\frac{\rho}{T})^{2/3}\int_{\omega_L}^{\omega_p} \log[n(\omega)]{\omega^{1/3}}d\omega$ with $n(\omega)=\frac{3T}{4\pi \rho}S_\eta(\omega)$. One can then compute the derivative of the entropy with respect to temperature to obtain heat capacity, at constant volume, $C_v \equiv {\theta}(\partial{\mathcal S}/\partial {\theta})$, while keeping constant all other variables. Figure~\ref{fig:S_dT} shows the evolution of $\partial {\mathcal S}/\partial {\theta}$ as a function of the effective temperature and a clear $\theta^{-1}$ power law is reported.  Indeed, injecting the Rayleigh-Jeans energy spectrum of Eq.~\eqref{eq:prediction_psd_k} into the above entropy definition gives $\partial {\mathcal S}/\partial \theta=k_B\theta^{-1}\int_{k_L}^{k_p}2\pi kdk\left(\frac{L}{2\pi}\right)^2$ (see dashed line in Fig.~\ref{fig:S_dT}) in good agreement with experiments. The effective heat capacity of the wave system is thus found to be independent of $\theta$ (see inset of Fig.~\ref{fig:S_dT}) as expected by the prediction
 \begin{equation}
 C^{\mathrm{Eq}}_v=\frac{1}{2}\mathcal{N}_f k_B \mathrm{,}
\label{eq:heat_cap} 
\end{equation}
where $\mathcal{N}_f = (k_p^2-k_L^2)L^2/(2\pi)$ is the number of~degrees of freedom. The heat capacity of large scales in equipartition thus corresponds to the wave energy per unit temperature stored in the band bounded by the system scale ($k_L$) and the forcing scale ($k_p$). When rescaled by $\mathcal{N}_f$, it becomes independent of the system and equals the Boltzmann constant $k_B$ as observed experimentally in the inset of Fig.~\ref{fig:S_dT}. As $k_p\gg k_L$, one has $\mathcal{N}_f\simeq 2\pi (L/\ell_p)^2 \sim 5600$. This number of degrees of freedom is huge compared to microscopic system ones. Still, it remains finite because the small-scale forcing provides a natural Fourier-space ultraviolet cutoff, which prevents the total energy divergence (that would be due to energy equipartition in an unbounded Fourier space). As here, the space is bounded by $k_p$, the equipartition and Rayleigh-Jeans spectrum thus emerge, due to nonlocal nonlinear wave interactions~\cite{zakharov1992}.

%Equation~\eqref{eq:heat_cap} is derived} by injecting the Rayleigh-Jeans energy spectrum of Eq.~\eqref{eq:prediction_psd_k} into the above entropy definition, leading to $\partial {\mathcal S}/\partial {\theta}=k_B{\theta}^{-1}\int_{k_L}^{k_p}{2\pi k}dk\left(\frac{L}{2\pi}\right)^{2}$. The prediction of Eq.~\eqref{eq:heat_cap} (dashed line) is in good agreement with experiments. 

%  large scales {in equipartition} of the wave system has a typical heat capacity value of $C_v \sim 10^{-22}$~J/K close to the prediction of Eq.\eqref{eq:heat_cap} of $20k_B\simeq 2.7\ 10^{-22}$~J/K, corresponding thus to the wave energy per unit temperature stored in the band bounded by the system size ($k_L$) and the forcing scale ($k_p$).
%{\color{blue}As a consequence, the related internal energy increases linearly with $T_{\mathrm{eff}}$ as by definition $dU=C_v dT_{\mathrm{eff}}$ and the large scales evolve accordingly to the Joule's first law.}
% Note that the decrease at low $T_{\mathrm{eff}}$ could be related to frozen out of the dof.

{\textit{Conclusions---}We have reported the first experimental evidence of a statistical equilibrium regime of large-scale hydroelastic tensional waves in coexistence with nonequilibrium small-scale dynamics dominated by random forcing. We also show that some thermodynamic concepts apply to describe the large scales of a turbulent system in a statistical equilibrium regime. In the future, we will investigate the dynamics of thermalization, namely how the large scales in statistical equilibrium decay over time when small-scale forcing is stopped. For such nonequilibrium systems with large scales in statistical equilibrium, it will be also of paramount interest to determine whether other classical tools of equilibrium or nonequilibrium statistical mechanics, such as the fluctuation-dissipation relations, fluctuation theorem, Green-Kubo relations, and large deviations theory, can be experimentally applied. These tools have recently been explored numerically in 3D turbulence~\cite{AlexakisEPL2023} and for modeling large scales of 2D geophysical turbulent flows~\cite{BouchetPR2012}. This approach could also be extended to other turbulent systems, such as capillary wave turbulence~\cite{michel2017}, elastic wave turbulence~\cite{miquel2021PRE}, optical wave turbulence~\cite{BaudinPRL2020}, superfluids and 3D Bose-Einstein condensates~\cite{ConnaughtonPRL2005}.  %quantum superfluid turbulence~\cite{},  
%(where forcing balances dissipation on average) 

%This approach could also be applied to various nonequilibrium systems, including active matter~\cite{Caprini2021}, electrical circuits~\cite{FalconPRE2009}, granular matter~\cite{AumaitreEPJB2001}, nonlinear optics~\cite{PicozziPR2014}, or quantum fluids~\cite{MicadeiPRL2021}.

%Further studies are still needed to explore if other classical tools of equilibrium statistical mechanics are experimentally valid in such out-of-equilibrium systems in statistical equilibrium, such as the fluctuation-dissipation laws, Green-Kubo relations, fluctuation theorem, and others, as explored numerically in hydrodynamics turbulence~\cite{AlexakisEPL2023}. 

 %Our results could be applied to physical, geophysical, or industrial turbulent flows where the large scales are in statistical equilibrium.
 
\begin{acknowledgments}
We thank M. Lanoy and F. Novkoski for fruitful discussions and Y. Le Goas and A. Di Palma for their technical support. This work was supported by the Simons Foundation MPS-WT-00651463 Project (U.S.) on Wave Turbulence and the French National Research Agency (ANR Sogood Project No.~ANR-21-CE30-0061-04 and ANR Lascaturb Project No.~ANR-23-CE30-0043-02).
\end{acknowledgments}

%\bibliographystyle{unsrt}
%\bibliography{ES_bib.bib}% Produces the bibliography via BibTeX.

\end{document}